\documentclass[pre]{revtex4}
 
\usepackage{graphicx}
\usepackage{dcolumn}
\usepackage{bm}

\begin{document}


\title{In the search for the low-complexity sequences in prokaryotic
and eukaryotic genomes: how to derive a coherent picture from global
and local entropy measures}

\author{Claudia Acquisti$^1$, Paolo Allegrini$^2$, Patrizia
Bogani$^1$, Marcello Buiatti$^1$,\\ Elena Catanese$^3$, Leone
Fronzoni$^{4,5}$, Paolo Grigolini$^{4,6,7}$, 
Giuseppe Mersi$^1$, Luigi Palatella$^4$}

\affiliation{ $^1$Dipartimento di Biologia Animale e Genetica
dell'Universit\`a degli Studi di Firenze, Via Romana 17, 50125
Firenze, Italy}

\affiliation{ $^{2}$Istituto di Linguistica Computazionale del
Consiglio Nazionale delle Ricerche, Area della Ricerca di Pisa, Via
Moruzzi 1, San Cataldo 56010 Ghezzano-Pisa, Italy}

\affiliation{$^{3}$Scuola Normale Superiore, Piazza dei Cavalieri, 56125
  Pisa, Italy}

\affiliation{ $^{3}$Dipartimento di Fisica dell'Universit\`{a} di Pisa
and INFM, via Buonarroti 2, 56127 Pisa, Italy}

\affiliation{$^{4}$Centro Interdipartimentale per lo Studio dei
  Sistemi Complessi, via S. Maria 28, 56126 Pisa, Italy}

\affiliation{ $^{5}$Center for Nonlinear Science, University of North Texas,
P.O. Box 311427, Denton, Texas 76203-1427 }

\affiliation{ $^{6}$Istituto dei Processi Chimico Fisici del CNR
Area della Ricerca di Pisa, Via G. Moruzzi 1,
56124 Pisa, Italy}
 
\begin{abstract}

We investigate on a possible way to connect the presence of Low-Complexity
Sequences (LCS) in DNA genomes and the nonstationary properties of base
correlations. Under the hypothesis that these variations signal a change in
the DNA function, we use a new technique, called Non-Stationarity Entropic
Index (NSEI) method, and we prove that this technique is an efficient way to
detect functional changes with respect to a random baseline. The remarkable
aspect is that NSEI does not imply any training data or fitting parameter, the
only arbitrarity being the choice of a marker in the sequence.  We make this
choice on the basis of biological information about LCS distributions in
genomes. We show that there exists a correlation between changing the amount
in LCS and the ratio of long- to short-range correlation.

\end{abstract}

\maketitle

\section{Introduction}

In the recent past there has been a significant interest in the search for
both long- and short-range correlation in DNA sequences
\cite{russi,voss,stanley}), The main results of these papers have been that
the amount of long-range correlation increases when moving from prokaryotes to
eukaryotes and from coding to non-coding sequences.  This paper is devoted to
discussing the same issue using two new theoretical tools that were not yet
available to the authors of the earlier papers. The former method, called
Diffusion Entropy (DE), was developed by the authors of
Refs.\cite{giacomo,nicola,PRE} for the purpose of defining the asymptotic
properties generated by the long-range correlation. When this method is
applied to DNA sequences, it affords global information, this being
unambiguous only in the ideal stationary case. The DNA sequences are non
stationary and the adoption of the DE method must be supplemented by a
non-stationarity indicator. This important indicator has been proposed in
Ref. \cite{giulia1}. We refer to it as Non-Stationarity Entropic Index (NSEI)
method. This is so because, as we shall see, this method is entropic in
nature, it vanishes in the stationary case, and it is larger the larger the
strength of non-stationarity. The measurement of long-range correlation is a
delicate issue. The detection of non-stationarity is another delicate
issue. The NSEI method addresses properly an even more delicate issue, this
being the non-stationary character of long-range correlation.

With the use of these two brand new techniques we prove that moving from
coding to non-coding and from eukaryotes to prokaryotes both the
non-stationarity degree and the amount of long-range correlation
increase. 
The deviations from stationarity and randomness are a sign of the
role played by selection during the process of life evolution. The same
argument applies to the increasing amount of non-stationarity, since this
phenomenon is thought to be associated with the selection-induced increasing
variety of molecular functions.

To make more solid this important biological conclusion we proceed as
follows. We take into proper account the ideas that have been developed in the
last few years to properly model the origin and the functions of the
constraints on the DNA sequences randomness \cite{ruffodna}.  More
specifically, we shed light into this issue by analyzing in the light of life
evolution some key Low-Complexity Sequences (LCS). 
These LCS are defined as sequences containing only AT
(homo-weak) or GC (homo-strong), or purines (GA), or pyrimidines (CT). The
reason for choosing these particular non-random sequences lies in the fact
that their presence is known to affect the DNA local conformational landscape,
giving rise to non-B structures (the B structure is the usual double helix),
and/or affecting the degree of bending, curving, twisting and rolling of DNA
molecules. Conformational features are known to be true 3-dimensional codes
for the interaction of DNA with proteins, a necessary condition for its
involvement in chromosome organization and in the key functions of
transcription and replication \cite{buiattiinpress}.

The outline of the paper is as follows: In section \ref{LCSsection} we
describe some basic properties of LCS distributions and their influence on DNA
structure. In section \ref{DEsection} we describe the global analysis trough
the DE of different genomes and compare it with LCS distribution data, while
in section \ref{Esisection} we illustrate the NSEI method. Section
\ref{Esiresultsection} aims at showing the results of two methods at work,
namely, the NSEI and a neural network trained by means of LCS lengths to
distinguish between coding and non-coding sequences. Finally Section
\ref{conclusion} is devoted to drawing some conclusions.

\subsection{Distribution of LCS 
and their putative functional meaning}\label{LCSsection}

Provata and Almirantis \cite{almirantis} analyzed the size distribution of
purine and pyrimidine clusters and found that coding and non-coding sequences
yield an exponential decay and a power law decay, respectively. We also note
that in eukaryotes the coding sequences relative amount is very low, whereas
it is the largely dominant fraction of prokaryotes genomes. Both properties
suggest to make an analysis of LCS length distributions in species placed in
different positions in the ``tree of life"\cite{buiattiinpress}. In a
preliminary paper \cite{buiatti2002} an algorithm was used to evaluate the LCS
density in a large number of bacterial, archaeal and eukaryotic genomes. A
clear difference was observed in the distributions of LCS relative amount
between coding and non-coding regions and between prokaryotes and
eukaryotes. In all genomes non-coding and Eukaryote sequences show the highest
amount of LCS. To understand whether high-density values were accompanied by a
random dispersion or by the presence of long LCS stretches, we decided to
carry out an analysis of LCS length distributions. Herein the attention is
centered on LCS known to give rise to non-B DNA conformations and to be
involved in the control of gene expression and recombination, namely the
homo-weak and the homo-strong.
In the case of AT-rich (homo-weak) sequences we found higher values of LCS
lengths in non-coding sequences vs. coding sequences and in eukariotes
vs. Prokaryotes.

\begin{figure}[!h]
\begin{center} 
\includegraphics[angle=0,width=14 cm]{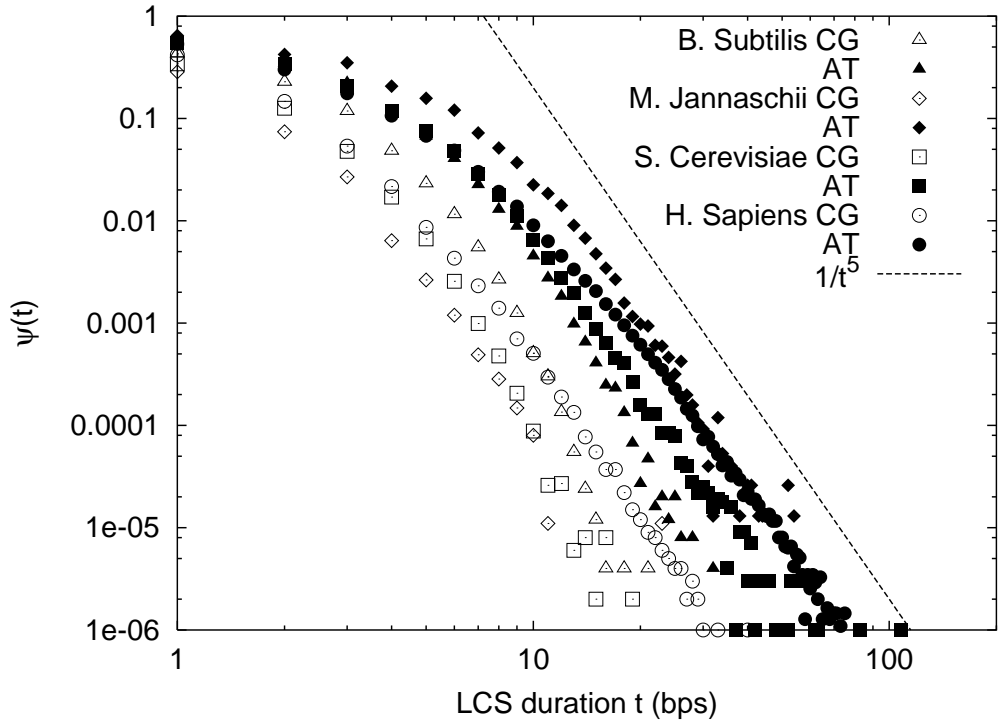}
\end{center}
\caption{\label{claudia} Probability distribution of lengths
of homo-weak (black symbols) and homo-strong (white symbols)
non-coding sequences. Notice that the length distributions
of homo-weak have a larger cutoff value; among these,
cutoff values are larger for eukaryotes, namely for S. Cerevisiae (squares)
and H. Sapiens (circles), than
for archaea M. Jannaschii (diamonds) and bacterium B. Subtilis
(triangles).
Dotted line represents an eye-guide decay $\propto t^{-5}$.}
\end{figure}

Fig. \ref{claudia} shows this property for homo-weak noncoding sequences in
some genomes, representative of archaea (M. Jannaschii), prokaryotes
(B. Subtilis), unicellular eukaryotes (S. Cerevisiae) and multicellular
eukaryotes (H. Sapiens, Chromosome I). We see an increasingly larger deviation
from a single exponential distribution going from prokaryotes to more complex
organisms with a larger amount of non-coding sequences. This property is less
trivial than it may appear, since it is true also for GC-rich (homo-strong)
sequences, as denoted by the curves with white symbols in
Fig. \ref{claudia}. All these curves show a clear deviation from Poissonian
behavior, even if their inverse-power-law decay has a large negative exponent
(about $-5$). However, for all genomes, the curves for homo-weak LCS have a
larger cut-off value than the corresponding ones for homo-strong LCS.
This fact is indicative of a selection pressure against the
lengthening of homo-strong sequence possibly to be attributed to
their rigidity and conformation landscapes different from those of
long AT-rich tracts.
This leads us to choose homo-weak occurrence as a marker in the correlation
analysis, as it will be explained below. Indeed, the presence of a decay
slower than Poissonian suggests the possibility that correlation with an
extend long range can be triggered by the presence of homo-weak and
homo-strong LCS.

Remarkably, the authors of \cite{pomodori} performed a thorough study of intra
and inter specific variability in upstream and coding sequences of a
series of key genes in Lycopersicum (tomato) (the gene for ACC
synthase), Nicotiana (the gene for Phytochrome), Primate species (the
genes EMX and OTX 2 involved in brain development).
All data showed a far higher variability in non-coding than in coding
sequences, preferentially localized in LCS. 
In other words, different varieties have
different lengths of AT-rich segments in the non-coding region before the
gene. Quantitative RT-PCR data (Reverse Transcription-Polymerase Chain
Reaction), a method to establish the amount of messenger-RNA, showed a clear
variation of gene expression corresponding to different lengths of the
homo-weak sequences.

Remarkably, the authors of \cite{pomodori} performed a thorough study of intra
and inter-specific variability in ACC-synthase upstream sequences. They
studied ACC-synthase upstream sequence in Nicotiana spp phytochrome regulatory
sequences and in the same regions for the primates genes EMX-2, OTX-2. The
results clearly showed that these LCS tend to be hyper-variables, i.e. they
show a mutation frequency significantly higher than other DNA stretches,
moving from and individual to another of the same species as well as from
species to species, coherently with the aforementioned data on recombination.
The interesting conclusion is that the three-dimensional code regulating the
gene expression may be associated to a strongly {\em dynamical} DNA structure,
with statistical constraints. This structure is probably associated to a
process with long-range correlation. In the following section we shall analyze
the long-range correlation present in the genome sequences and the relation
between this property and the presence of LCS.

\section{The diffusion entropy method 
applied to homogeneous sequences}\label{DEsection}

\subsection{Global Diffusion Entropy}

In recent papers \cite{giacomo,nicola,PRE} a new way of revealing long-range
correlation, based on the detection of anomalous scaling has been proposed.
In short, one defines a ``marker'' on a time sequences, and evaluates the
probability $p(x;t)$ of having the number $x$ of markers in a window of length
$t$. The evaluation of $p(x;t)$ is done by moving a window of length $t$ along
the sequences and counting how many times one finds $x$ markers inside this
window. At this stage $p(x;t)$ is obtained by dividing this number by the
total number of windows of size $t$, which is obviously $N-t+1$, where $N$ is
the total length of the sequence. In the case of large values of $x$ and $t$,
the continuous approximation is legitimate, and, in the ergodic and stationary
condition, a scaling relation is expected, namely In case of large values of
$x$ and $t$, the continuous approximation is legitimate, and, in the ergodic
and stationary condition, we expect scaling, namely, we expect
\begin{equation}
\label{scaling}
p(x;t)=\frac{1}{t^{\delta}} F \left( \frac{x}{t^{\delta}} \right),
\end{equation}
where $\delta$ is said to be the {\em scaling index} and $F$ is a
function, sometimes called ``master curve''.  If $F$ is the Gauss
function, $\delta$ is the known Hurst index, and if the further
condition $\delta=0.5$ is obeyed, then the process is said to be
Poissonian, namely there is no long-range memory regulating the
occurrence of markers in time.
It is straightforward to show that the Shannon Information

\begin{equation}
\label{shannon}
S(t)=\int_{-\infty}^{\infty}dx p(x;t) \ln p(x;t)
\end{equation}
with conditiom (\ref{scaling}) leads to

\begin{equation}
S(t)=k + \delta \ln t,
\end{equation}
where $k$ is a constant. A linear fit of $S(t)$ in log-normal paper
allows an extimation of $\delta$.

\subsection{The CMM model of DNA}

In recent years many groups have agreed to consider DNA sequences as a mixture
of long- and short- range correlation, the latter blurring the strength of the
former. A simple model, originally proposed by Araujo et al. \cite{araujo},
and later adopted independently by the authors of Refs. \cite{maria},
\cite{marcobuiatti2} and \cite{zanettedna}, to name a few, is the Copying
Mistakes Map (CMM)\cite{maria}. This is the superposition of two models, one
consisting of a pure random choice, and the other of an intermittent generator
of homogenous sequences, yielding long-range correlation. Both models provide
a sequence, where the $i$-t nucleotide of the resulting DNA sequence is taken
from the first component (the white one) with probability $1-\epsilon$ and
from the second with probability $\epsilon$. The resulting correlation
function is proven \cite{maria} to be
\begin{equation}
\label{correlationfunction}
C(t)= \delta_{i,0} + \epsilon^{2} {\left(\frac{T}{T+t}\right)}^{\beta},
\end{equation}
where $\delta_{i,j}$ is the Kroeneker delta, while $T$ and $\beta$ are
two positive parameters of the intermittent model. Long-range memory
is normally characterized by the condition $\beta<1$, which makes the
correlation function (\ref{correlationfunction}) not integrable.

\begin{figure}[!h]
\begin{center}
\includegraphics[angle=0,width=12 cm]{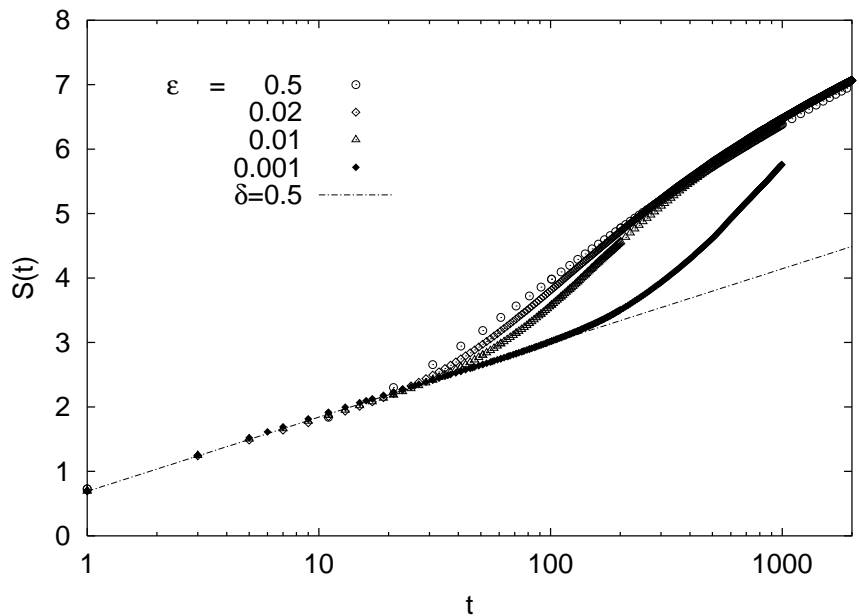}
\end{center}
\caption{\label{figTeo} Diffusion Entropy for CMM's
with the same $\delta$, and various values of $\epsilon$.}
\end{figure}

As shown in ref. \cite{maria}, the second moment and the Hurst analysis reveal
a short-time behavior which is dominated by the random component, while the
correlated part dominates in the long-time limit. By the same token, DE method
applied to CMM sequences, yields for $S(t)$ a curve that starts with a slope
of $\delta=0.5$, and then after a knee, tends asymptotically to higher value
of $\delta$ corresponding to the L\'evy scaling $\delta=1/(\beta+1)$. This
behavior, as reported in Ref.~\cite{latora}, is illustrated in
fig.\ref{figTeo}, which shows that the position of the knee is a monotonic
function with respect to $\epsilon$: the larger $\epsilon$ the smaller the
position of the knee.

\subsection{DE at work on real genomes}

\begin{figure}[!h]
\vspace{1cm}
\begin{center}
\includegraphics[angle=0,width=10 cm]{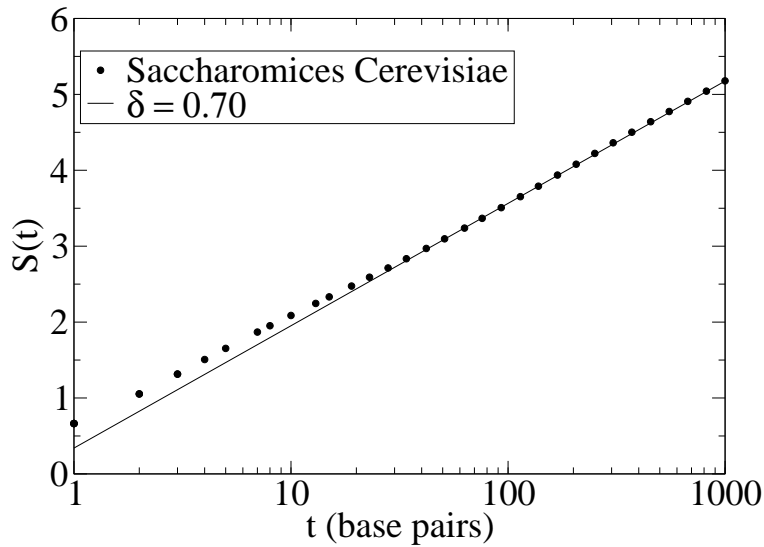}
\end{center}
\caption{\label{saccharo_DE} Diffusion Entropy analysis for {\em
    Saccharomices Cerevisiae}. }
\end{figure}

As earlier shown, the DE method establishes the values of two quantities: the
ratio of the correlated to the uncorrelated component intensity 
of the CMM model
(through the position of the knee) and the anomalous scaling index $\delta$.
As an example, we show in fig. \ref{saccharo_DE} the DE analysis on the
chromosome IV of {\em Saccharomices Cerevisiae}.  In Tab.\ref{tab1} we report
the results corresponding to different genomes grouped according to whether
they are bacteria, archaea or eukaryotes. All these analyses were performed
using as marker the homo-weak occurrence in the series. This means that we put
$1$ at position $i$ if the $i$-th nucleotide {\em and} the previous one are
weak bases (AT), and $0$ elsewhere. This choice was suggested by the
non-exponential decay in the length distribution of homo-weak sequences shown
in fig.\ref{claudia}a. We made the hypothesis that this non-exponential
behavior can be connected with the long-range correlation observed in the
genomes.

\begin{table}[h]\label{DEDNA}
\begin{tabular}{|c|c|c|}
\hline
{\bf Genomes} & {\bf knee} $\pm 5$ & $\delta \pm .04$\\
\hline
\multicolumn{3}{|c|}{{\bf Bacteria}} \\
\hline
M. pneumoniae * & 80 & 0.76 \\
T. maritima * & 90 & 0.75 \\
R. prowazekii *  & 90 & 0.60 \\
H. influenzae *   & 60 & 0.71 \\
B. subtilis *  & 100 & 0.78 \\
E. coli *  & 100 & 0.73 \\
A. aeolicus *  & 90 & 0.72 \\
Synechocystis Sp. *  & 80 & 0.73 \\
\hline
\multicolumn{3}{|c|}{{\bf Archaea}} \\
\hline
M. thermoautotrophycum *  & 60 & 0.76 \\
A. fulgidus *  & 60 & 0.68 \\
M. jannaschii *  & 40 & 0.68 \\
P. abyssii *  & 100 & 0.76 \\
\hline
\multicolumn{3}{|c|}{{\bf Eukaryotes}} \\
\hline
S. cerevisiae *  & 50 & 0.70 \\
H. sapiens Chr. 1 *  & 20 & 0.88 \\
C. elegans *   & 5 & 0.76 \\
A. thaliana *  & 15 & 0.75 \\
\hline
\end{tabular}
\caption{\label{tab1}  The * denotes complete genomes or chromosomes. The analysis was performed
using the homo-weak occurrence as a marker.}
\end{table}

Tab.\ref{tab1} shows two interesting results: the former is that the value of
$\delta$ does not seem to depend on the position in the evolutionary tree. The
latter is that the position of the knee seems to change with moving from
bacteria to archaea and eukaryotes. As shown in the table, the weight of the
correlated component in bacteria is larger than in eukaryotes, archaea being
somewhat intermediate between bacteria and eukaryotes. It is also worth
noticing that within eukaryotes, the unicellular, almost intronless,
S. Cerevisiae has the largest knee value.

\section{The Non-Stationarity Entropic Index}\label{Esisection}

In order to study the local properties of DNA sequences with special focus on
the distribution and variability of homogeneous sequences, we use a new
method, the NSEI method. This method is derived from the DE method
\cite{giulia1} as an earlier method, called CASSANDRA algorithm
\cite{giulia1,granada}.  The DE method affords global information, while both
CASSANDRA and NSEI method aim at affording local information. The CASSANDRA
method measures the rate of transition from dynamics to thermodynamics through
comparison with ordinary Brownian motion. The NSEI method focuses on the local
deviation from the stationary behavior.  The first step of the NSEI method is
to build the diffusion entropy function $S(t)$ for each window of size $L$
with $L \gg t$, thought of a complete sequence. In this way we obtain several
functions $S_j(t)$, where the subscript $j$ refers to the portion of the
sequence used to calculate $S_j(t)$. Formally, we write
\begin{equation}
x_j(t,l) \equiv \sum \limits_{i=j+l}^{j+l+t} \xi_i \:\:  ,
\mathrm{with} \:\: 0<l<L-t
\end{equation}
where $\xi_i$ is the DNA sequences after marking the homogeneous sequences, as
earlier ilustrated, and $j$ and $L$ denote the beginning and the length of the
big window, respectively. We assign to the integers $j$ multiple values of a
given integer $J$, namely $j=0,J,2J,\cdots$, thus avoiding too large overlaps
between two windows of length L, corresponding to two consecutive choices of
$j$. At fixed $t$ and $L$ we obtain for different $l$ a probability
distribution $p_j(x,t)$. Calculating the entropy of this distribution, we
obtain
\begin{equation}
S_j(t) \equiv \int {\rm d}x p_j(x,t) \log p_j(x,t)
\end{equation}
Now we are ready to define the non- stationarity index, 
indicated as $\Xi_j$, as follows

\begin{equation}
\Xi_j= \int\limits_0^t {\rm d}t' \left ( S_j(t')-\bar S_j(t')  \right )
\end{equation}
where
\begin{equation}\label{media}
\bar S_j(t) \equiv \frac{1}{m} \sum \limits_{i=j-m-1}^{j-1} S_i(t)
\end{equation}
Note that $m$ is the number of windows involved by the average process
of Eq.(\ref{media}). Depending on the step $J$ by which $j$
increases we obtain that the comparison between local and previous
diffusion behavior involves $m \cdot J $ points of the sequence.
Note that if the sequence is perfectly stationary we have $\Xi_j=0$,
otherwise we have a value different from zero.

\section{DNA function and local correlation}\label{Esiresultsection}

\begin{figure}[!h]
\begin{center}
\includegraphics[angle=0,width=12 cm]{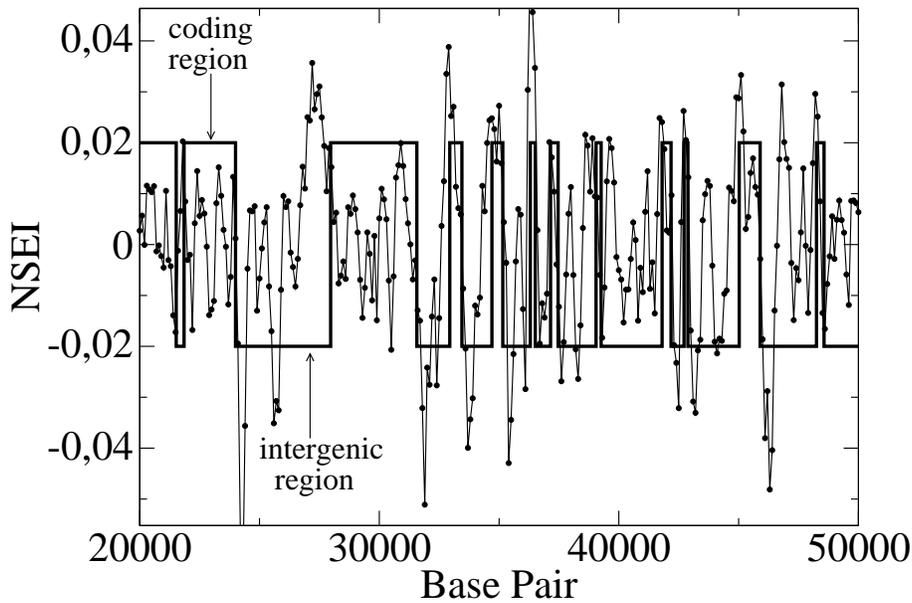}
\end{center}
\caption{\label{figurasaccharo} Non-Stationarity Entropic Index for {\em
    Saccharomices Cerevisiae}. The abcissa axis denote the position of
    center of the large window with length $L=400$.  Other parameters
    of the analysis are: the jump size $J=100$, the number of averaged
    windows $m=4$ and maximum time $t=12$ for entropy evaluation. The
    thick line denotes the displacements of coding and noncoding
    portion of the genome, }
\end{figure}

In Section \ref{LCSsection} we saw that non-coding sequences contain a number
of homogeneous sequences larger than coding sequences. Experimental evidence
shows that LCS tend to occur either in the upstream region of each gene,
i.e. immediately before the occurrence of a gene, or in introns. It is also
known that both upstream sequences and, to a lesser extent, introns, are the
regions endowed with regulatory roles in which LCS are involved. In this
section we evaluate the change of local correlation, according to the NSEI
method, and we see that this change is often associated with 
the beginning of a coding
portion. Thus, using biological consideration, we are led to establish a
plausible link between the change of DNA sequence correlation and the
functional change of the sequence. In other words, we expect that the boundary
between a coding and a non-coding segment is revealed by either a positive or
negative peak of the NSEI index, as a function of the position $j$ along the
sequence. In Fig. \ref{figurasaccharo} we show an example where this
hypothesis is verified to a satisfactory extent. As earlier stated, the
analysis was carried out using as marker the occurrence of homo-weak
sequences. Peak recognition has been realized in the following way.  The data
have been smoothed through a moving linear interpolation with ten points. The
maximum indexes were identified looking at the ``derivative'' thus obtained;
only maximum indices with a value larger than the standard deviation,
calculated over the whole signal, were selected.

The NSEI method should reveal the beginning and the end of a LCS with
peaks. It is important to point out that the NSEI method, although being the
most convenient indicator of non-stationarity, within the theoretical
framework of DE method, does not have a perfect efficiency.  This means that
the NSEI signals correctly the existence of a boundary between two distinct
regions. This is a true positive, denoted by the symbol $TP$. There are cases,
however, when the NSEI signals a boundary that does not exist. This is called
a false positive, and it is denoted by the symbol $FP$. By the same token we
denote by $TN$ and $FN$, the true and false case, respectively, of a missing
boundary. The total number of borders is given by $FN+TP$, while the total
number of cases signaled by the NSEI, true or false, is given by $TP+FP$.  It
is worth noticing that the observed peaks, in addition to corresponding to
true peaks, must be found in position close enough to the position of the real
borders, namely at a distance smaller than a critical value $T$, which is of
the same order as $L$.

We are now in the proper position to establish the accuracy of the NSEI. We
can judge this method satisfactory if we prove that it affords a number of
correct guesses significantly larger than the random choice.  To assess this
issue we apply to the NSEI method a technique of assessment frequently used in
Information Extraction \cite{precall}. We define two quantities, Precision and
Recall. Precision $P$ is a measure of how accurately the displacements of
coding/non-coding boundaries are guessed, while Recall $R$ is a measure of how
many of these regions are correctly located.Formally, we write
\begin{equation}
P=\frac{TP}{TP+FP}
\end{equation}
and

\begin{equation}
R=\frac{TP}{TP+FN}
\end{equation}

We compare the $P$ and $R$ of our predictor to a baseline score associated
with $TP+FP$ random choices for the position of the boundaries. The ``true''
classification was done according to documents accompanying DNA sequences in
Genbank database, regardless whether the annotation was experimental or
putative. Although this makes our esteem or $P$ and $R$ questionable, we think
that the improvement with respect to the random baseline will not diminished
after a more accurate experimental annotation. Results are summarized in table
\ref{presigionrecolle}.  As shown in the table, the NSEI method reveals
statistically significant improvement with respect to the baseline, especially
for eukaryotes, where the marker adopted (homo-weak) has a larger biological
significance.

\begin{table}
\begin{tabular}{|c|cccc|c|c|c|c|}
\hline
Genome & $TP+FP$ & $TP$ & $FP$ & $TP+FN$ & $P$ & $R$ & Random $P$ &
Random $R$ \\
\hline
\hline
A. aeolicus & 1089 & 457 & 632 & 1744 & 0.420 & 0.262 & 0.278 & 0.173 \\
A. fulgidus & 1497 & 879 & 618 & 2992 & 0.587 & 0.294 & 0.326 & 0.163 \\
B. subtilis & 3157 & 1673 & 1484 & 7584 & 0.530 & 0.221 & 0.424 & 0.177 \\
B. burgdorferi & 627 & 237 & 390 & 1252 & 0.378 & 0.189 & 0.314 & 0.157 \\
C. tetani & 2517 & 1236 & 1281 & 6556 & 0.491 & 0.189 & 0.426 & 0.163 \\
V. cholerae & 2898 & 1631 & 1267 & 6166 & 0.563 & 0.265 & 0.381 & 0.179 \\
E.coli & 3235 & 1966 & 1269 & 6894 & 0.608 & 0.285 & 0.387 & 0.181 \\
H. influenzae & 1215 & 619 & 596 & 2920 & 0.509 & 0.212 & 0.401 & 0.167 \\
H. pylorii & 1143 & 551 & 592 & 2406 & 0.482 & 0.229 & 0.349 & 0.166 \\
M. pneumoniae & 504 & 241 & 263 & 926 & 0.478 & 0.260 & 0.313 & 0.170 \\
N. ghonoreae & 1607 & 1065 & 542 & 3668 & 0.663 & 0.290 & 0.428 & 0.187 \\
P. aeruginosa & 4037 & 2796 & 1241 & 9310 & 0.693 & 0.300 & 0.381 & 0.165 \\
P. abysii & 1216 & 638 & 578 & 2302 & 0.525 & 0.277 & 0.323 & 0.170 \\
R. prowazekii & 762 & 301 & 461 & 1439 & 0.395 & 0.209 & 0.372 & 0.197 \\
S. typhi & 3312 & 1984 & 1328 & 7260 & 0.599 & 0.273 & 0.390 & 0.178 \\
Synechocistis spp. & 2464 & 1415 & 1049 & 5729 & 0.574 & 0.247 & 0.425 & 0.183 \\
T. maritima & 1309 & 550 & 759 & 2154 & 0.420 & 0.255 & 0.264 & 0.160 \\
D. radiodurans & 2169 & 1302 & 867 & 4531 & 0.600 & 0.287 & 0.376 & 0.180 \\
\hline
\hline
M. Jannaschii & 1088 & 578 & 510 & 2858 & 0.531 & 0.202 & 0.424 & 0.161 \\
M. thermoautotrophicum & 1305 & 893 & 412 & 3014 & 0.684 & 0.296 & 0.407 & 0.176 \\
\hline
\hline
S. cerevisiae & 6733 & 4096 & 2637 & 10506 & 0.608 & 0.390 & 0.337 & 0.216 \\
\hline
\hline
A. thaliana & 7698 & 2203 & 5495 & 4201 & 0.286 & 0.524 & 0.137 & 0.251 \\
H. sapiens Chr1 & 7413 & 86 & 7327 & 208 & 0.012 & 0.413 & 0.007 & 0.242 \\
D. melanogaster & 7430 & 1901 & 5529 & 2577 & 0.256 & 0.738 & 0.083 & 0.239 \\
C. elegans & 7007 & 1175 & 5832 & 4119 & 0.168 & 0.285 & 0.131 & 0.223 \\
\hline
\end{tabular}
\caption{\label{presigionrecolle}The parameters for the analysis
    reported herein are: large windows with length $L=400$, jump size
    $J=100$, number of averaged windows $m=4$ and $t=12$ for maximum
    time of entropy evaluation. Different boxes divide species into
    four groups: bacteria, archaea, unicellular and multicellular eukariotes.}
\end{table}

We end this section by reporting results stemming from a completely different
way of detecting coding and non-coding sequences. This alternative way
confirms the correlation between the changes of the amount of LCS and entropy
changes, when moving from the coding to the non coding regions. A three layer
feed forward neural network was built and trained on homogeneous sequences
(homo AT; homo GC; homopurine; homopyrimidine) \cite{neuralnets}. The network
involves eight neurons in the input layer, four in the intermediate, one in
the output one. The network has been trained on 100 tracks each of 300
nucleotides of known coding and on 100 for known non coding regions. Inputs of
the network are 1) percentage of the sequence covered by homogeneous tracks 2)
average length of homogeneous tracks in the observed sequence. Then 900 tracks
have been screened with the trained network for each class. This operation has
been performed for each species listed in the table \ref{reteneurale}. The
precision level of annotation (identification of the two classes) is very high
as seen from Q and W values. For true positives (Q), moreover, eukaryotes seem
to perform better than prokaryotes, coherently with the higher
non-coding/coding ratio.

It is worth noticing that this second way of detecting the change of the DNA
function is complementary to the first one. Indeed the NSEI method, which does
not require any form of training, yields a model-oriented location of 
the function change on a scale comparable with $L$. On the other hand the 
neural network technique is more precise and can become very efficient 
when performed on the basis of evidence obtained by means with NSEI.
\begin{table}
\begin{tabular}{|c|ccccccc|}
\hline
Genome & $Ntot$ & $TP$ & $ncp$ & $TN$ & $ncn$ & $Q$ & $W$ \\
\hline
\hline
A. aeolicus & 900 & 477 & 110 & 658 & 97 & 0.767 & 0.678 \\
M. Jannaschii & 900 & 730 & 46 & 531 & 95 & 0.727 & 0.811 \\
\hline \hline
S. cerevisiae (chr 1) & 900 & 566 & 79 & 813 & 40 & 0.923 & 0.761 \\
S. cerevisiae (chr 10) & 900 & 651 & 45 & 700 & 41 & 0.804 & 0.774 \\
\hline \hline
A. thaliana (chr 1)& 900 & 702 & 20 & 773 & 26 & 0.874 & 0.813 \\
C. elegans (chr 1) & 900 & 853 & 5 & 846 & 7 & 0.948 & 0.953 \\
D. melanogaster & 900 & 627 & 17 & 889 & 6 & 0.992 & 0.776 \\
\hline
\end{tabular}
\caption{\label{reteneurale}The parameters fo the analysis
    reported herein are:  $TP$=true positive non coding tracks,
    $TN$=true positive coding tracks, $ncp$ = non classified non coding tracks,
$ncn$ = non classified coding tracks
,$FP$ = coding tracks classified as non coding,
$FN$ = non coding tracks classified as coding,
$Q = TP/(TP+FP)$ and
$W = TN/(TN+FN)$. Different boxes divide bacteria, yeasts and
    multicellular eukaryotes.
}
\end{table} 


\section{Conclusions}\label{conclusion}

In this paper we discussed how to derive a coherent picture from the study of
LCS, local properties, and long-range correlation, global properties. Thanks
to the DE technique we showed that long-range correlation seems to have a
quite homogeneous behavior when looked at using a large scale. Variations in
the transient behavior, however, confirm an increase of the non-random
component with moving from bacteria to archaea, from archaea to the
unicellular eukaryote S.  cerevisiae, and from this to multi-cellular plants
and animals. This increase throughout the evolution scale is consistent with
that of non-coding sequences, which are known to cover less than 10\% of
prokaryotes genomes and well over 90\% of the eukaryotes ones. A plausible
hypothesis, to explain these properties in terms of function, is that
non-coding sequences can be more freely ``filled" with ``hidden codes".  These
hidden codes are mainly used for the recognition of protein molecules,
necessary for the organization and function of the DNA sequence and other
ligands.  Remarkably, these codes, and in general non-coding sequences as
well, tend to be hyper-variable, due to the presence of LCS, suggesting {\em a
positive role of mutational noise in regulatory sequences} (see Ref
\cite{buiattiinpress} for a through discussion).

To afford some more details on why to connect LCS,
conformational landscapes, regulation and structure of genomes, we
should note that some constraints may play a role in 
eukaryotes for DNA ``packaging" in
nucleosomes and chromosomes. Moreover, it is well known that DNA activation
for transcription and replication is due to the recognition by proteins of DNA
regions, and vice-versa, based on complementary conformations leading to
formation of complexes. Protein-DNA complexes involve at the same time a high
number of molecules where interaction needs a specific global organization
favored by DNA curving, bending, twisting and rolling at the right
points. This may require the presence of non-random tracts with the needed
conformational landscape. 
The strategy reported herein, namely the
adoption of NSEI and of neural networks, fulfills
the biologist's need for research leading to the characterization 
of the living state of
matter (LSM) \cite{buiabuia} on one hand, to the identification of specific
functional sequences in sequenced genomes on the other.

We investigated on the connection between correlation and LCS, and between LCS
and the regulation of a gene transcription.  We showed that the variation in
the statistics of homo-weak sequences, i.e. a particular kind of LCS,
yielding specific DNA conformational
landscapes, 
strongly
correlates with changes in DNA function and, especially, that these sequences
signal the beginning of a coding region. We studied these variations with the
help of the NSEI algorithm obtaining satisfactory results. 
The accuracy achieved, although affording significant
information, is not high, in comparison with other annotation methods
currently used by molecular biologists. In fact, with some of these methods,
the percentage of failure is now around 10\%-15\%. Better results were herein
obtained with a neural network trained on LCS, the class of non-random
sequences. 
The NSEI method, however, a model-based technique, affords a way to
scan the whole genome, in the search for non-stationary segments,
where a transition from a given degree of correlation to another
occurs: The putative boundaries between coding and non-coding DNA may
generate the segmented input tracks for the neural network, a method
with a larger accuracy for the final categorization.
The NSEI method is based on a DNA dynamic model, compatible 
with the DE method, which signals the long-range corelation 
of the sequence. If the correlation strength undergoes 
local changes, these are signaled by the NSEI method.
Moreover,
the increase in complexity in the ``tree of life'' correlates
with an increase in non-coding DNA and in the number
of coding/non-coding boundaries. 
A high precision on the same task of recognizing coding sequences
has been obtained by other authors with algorithms based on specific motifs
known to be present in non-coding regions or exploiting existing periodicities
\cite{herzelgrosse2000}. Probably an even higher precision could be obtained
through the combination of all methods. It is therefore quite important that
the NSEI results were obtained without using all these properties but only the
variation in the statistics of homo-weak sequences.
 
A reasonable interpretation of these data all taken together, could be that
several kinds of functionally constrained sequences have been fixed in genomes
throughout evolution and that average higher correlation values in non-coding
sequences are a result of this process. However at a local level, while to
some extent LCS can be present also in coding regions, precise annotation
(distinction between coding and non-coding sequences) has to rely on the
detection of constraints (codes) specific for each class of DNA regions.  For
instance, while a 3-periodical behavior, in phase with the triplet
``universal" code is typical of coding regions \cite{maria,herzelgrosse2000},
specific motifs and rather long LCS have been shown to be distinctive of
non-coding regions, which are free from coding constraints \cite{maria}. In
other words, the relatively low level of constraints corresponding to
long-range correlations (and to the presence of LCS) in transcribed and
translated sequences can be attributed to the fact that in that case selection
has not been acting on DNA, but on proteins, and there is not a direct
relationship between constraints in the two classes of molecules.

As a final remark, looking at LCS, we point out an almost paradoxical result,
namely that hypervariable sequences carry a constraint. LCS hypervariability
is known to be favored in regions of genes where variation, leading to protein
variability, is essential. For instance, it can be useful for the right
response to hypervariable pathogens. Lower complexity levels in eukaryotes
seem therefore to derive from a wide series of constraints ranging from
periodicities to different kinds of short-range and long-range correlations
and to the presence of LCS of varying length. This hypothesis is supported by
the present work, suggesting the NSEI as a tool for analyzing data,
when looking at this kind of constraints, because of its locality and its wise
use of statistics. Furthermore, we want to point out that the NSEI emerges
from the theoretical background of Ref. \cite{giulia1}. The DE method was
originally introduced as an efficient technique to evaluate scaling, this
being a global property, implying stationary condition. 
The NSEI is the best
way, known to us in this moment, to address the challenging issue of
non-stationarity from the same dynamic approach to complexity as that behind
the foundation itself of the DE method.
From a conceptual point of
view, more than from an application point of view, we want to stress
that this technique of analysis, resting on entropy formalism, might
have the important role of helping the foundation of the LSM theory
\cite{buiabuia,gerardo}. As explained in Ref. \cite{buiabuia}, in fact, 
the relation between
hyper-variable sequences and regulation is a key aspect of the LSM
perspective, and the investigation of this issue based on the NSEI
might make easier for the advocates of the LSM perspective to express
the main ideas of this theoretical proposal with the language of
statistical mechanics, probably anomalous statistical mechanics.

\emph{Acknowledgments} PG gratefullly acknowledges financial support from ARO
through Grant DAAD19-02-0037


\begin{thebibliography}{99}

\bibitem{russi} Li, W. and  Kaneko, K. 1992. Long-range correlation and
partial 1/f spectrum in a noncoding DNA sequence. Europhys. Lett., 17,
655--660.

\bibitem{voss} R. Voss. Evolution of long-range fractal correlations
and 1/f noise in DNA base sequences. Phys. Rev. Letters,
68:3805--3808, 1992.

\bibitem{stanley} C. -K. Peng, S. V. Buldyrev, S. Havlin, M. Simons,
H. E. Stanley, and A. L. Goldberger, Phys. Rev. E, {\bf 49}, 1685
(1994); C. -K. Peng, S. Havlin, H. E. Stanley, and A. L. Goldberger,
Chaos, {\bf 5}, 82 (1995);

\bibitem{giacomo} P. Grigolini, L. Palatella, G. Raffaelli, Asymmetric
Anomalous Diffusion, an Efficient Way to Detect Memory in Time
Series, Fractals 9, 439-449 (2001).

\bibitem{nicola}  Nicola Scafetta, Patti Hamilton and Paolo Grigolini, 
The Thermodynamics of Social Process: the Teen Birth Phenomenon,
Fractals, 9, 193 (2001); Nicola Scafetta, and Paolo Grigolini,
Scaling detection in time series: diffusion entropy analysis,
Phys. Rev. E 66, 036130 (2002)

\bibitem{PRE} P. Allegrini, P. Grigolini, P. Hamilton, L. Palatella,
G. Raffaelli, Memory beyond Memory in Heart Beating, a Sign of Healthy
Physiological Condition Phys. Rev. E 65, 041926 (2002).

\bibitem{giulia1} P. Allegrini, V. Benci, P. Grigolini, P. Hamilton,
M. Ignaccolo, G. Menconi, L. Palatella, G. Raffaelli, N. Scafetta,
M. Virgilio, J. Yang, Compression and Diffusion: a Joint Approach to
Detect Complexity, Chaos Solitons \& Fractals 15, 517-535 (2003).

\bibitem{ruffodna} P. Li\'o, S. Ruffo, M. Buiatti, Third Codon G + C
Periodicity as a Possible Signal for an "Internal" Selective
Constraint, Journal of Theoretical Biology, 171(2),215-223 (Nov 21,
1994).
\bibitem{buiattiinpress} M. Buiatti, P. Bogani, C. Acquisti, G. Mersi, 
L. Fronzoni, The living state of matter between noise and homeorretic
constraints, in the forthcoming volume 
{\em Interdisciplinary applications of ideas from 
nonextensive statistical mechanics and thermodynamics} 
ed. by Murray Gell-Mann and Constantino Tsallis,
Oxford University Press, Oxford (2003). 

\bibitem{almirantis} A. Provata and Y. Almirantis, 
Scaling properties of coding and non-coding DNA sequences, 
Phys. A, v. 247 p.p. 482 (1997). 

\bibitem{buiatti2002} M. Buiatti, C. Acquisti, G. Mersi, P. Bogani, 
The biological meaning of DNA correlations, in Biology and Medicine,
G. Losa, Birkhauser, Basel p. 235 (2002).

\bibitem{pomodori} Bogani, P., Simoni, A., Li\'o, P., Germinario, A.,
Buiatti, M.  (2001). Molecular variation in plant cell populations
evolving in vitro in different physiological contexts. Genome,
44:1-10; Bogani P., Scialpi, A., Masieri M., Nardi M., Rosati A., Gori
M., Buiatti M. (2003).  Functional markers for the study of genetic
variation in tomato. Proc. 7th Int. Con. Plant. Mol .Biol., in press.

\bibitem{araujo} S.V. Buldyrev, A.L. Goldberger, S. Havlin,
C.-K. Peng, M. Simons and H.E. Stanley , Generalized L\'evy walk model
for DNA nucleotide sequences. Phys. Rev. E 47 (1993), pp. 4514-4523

\bibitem{maria} P. Allegrini, M. Barbi, P. Grigolini and B. J. West,
Dynamical model for DNA sequences, Phys. Rev. E, v. 52, p. 5281
(1995).

\bibitem{marcobuiatti2} P. Allegrini, M. Buiatti, P. Grigolini and
B. J. West, Non-Gaussian statistics of anomalous diffusion: The DNA
sequences of prokaryotes, Phys. Rev. E, v. 58, p. 3640 (1998).

\bibitem{zanettedna} D. E. Strier, D. H. Zanette, Self similarity in a
model of genetic microevolution, Physica A, Volume 257, 1998, Pages
530-535

\bibitem{latora} Nicola Scafetta, Vito Latora and Paolo Grigolini,
L\'evy statistics in coding and non-coding nucleotide sequences,
Physics Letters A 299 (5-6), 565-570 (2002);
Nicola Scafetta, Vito Latora and Paolo Grigolini,
Scaling without detrending: the diffusion entropy method applied to 
the DNA sequences, Phys. Rev. E 66, 031906 (2002).


\bibitem{granada} P. Allegrini, P. Grigolini, L. Palatella,
G. Raffaelli, M. Virgilio, Facing non-stationarity Conditions with a
New Indicator of Entropy Increase: the CASSANDRA Algorithm, in: Novak,
M.N. (ed.): Emergent Nature. World Scientific, Singapore (2002)
173-184.

\bibitem{precall} See, for instance, K.S. Jones. Information
Retrieval Experiment. Butterworth and Co., 1981.

 
\bibitem{neuralnets} D. Graupel, Principle of Artifical Neural
Networks, World Scientific, Singapore (1997).

\bibitem{herzelgrosse2000} I. Grosse, H. Herzel, S.V.
Buldyrev, H.E. Stanley , Species indipendent of mutual 
information in coding and noncoding DNA, Phys. Rev. E, v.
61, p. p. 5624 (2000).

\bibitem{buiabuia} M. Buiatti, N. Buiatti, Towards characterization 
of the living state od matter, Chaos, Solitons and Fractals, 
these Procedings.

\bibitem{gerardo} P. Allegrini, G. Aquino, P. Grigolini, L. Palatella, 
A. Rosa, {\em Breakdown of the Onsager Principle as a Sign of Aging},
submitted to Phys. Rev. E.
 
\end{thebibliography}
\end{document}